\documentclass[9pt, conference]{IEEEtran}
\IEEEoverridecommandlockouts

\usepackage{cite}
\usepackage{amsmath,amssymb,amsfonts}
\usepackage{algorithmic}
\usepackage{array}
\usepackage{graphicx}
\usepackage{textcomp}
\usepackage{xcolor}
\usepackage{booktabs}
\usepackage{bm}
\usepackage{ragged2e}
\usepackage[hidelinks]{hyperref}
\usepackage{pifont}
\newcommand{\cmark}{\ding{51}}

\begin{document}

\title{PDMX: A Large-Scale \textbf{P}ublic \textbf{D}omain \textbf{M}usic\textbf{X}ML Dataset for Symbolic Music Processing
\thanks{© 2025 IEEE.  Personal use of this material is permitted. Permission from IEEE must be obtained for all other uses, in any current or future media, including reprinting/republishing this material for advertising or promotional purposes, creating new collective works, for resale or redistribution to servers or lists, or reuse of any copyrighted component of this work in other works.}
}

\author{\IEEEauthorblockN{Phillip Long}
\IEEEauthorblockA{\textit{Computer Science \& Engineering} \\
\textit{UC San Diego}\\
La Jolla, USA \\
p1long@ucsd.edu}
\and
\IEEEauthorblockN{Zachary Novack}
\IEEEauthorblockA{\textit{Computer Science \& Engineering} \\
\textit{UC San Diego}\\
La Jolla, USA \\
znovack@ucsd.edu}
\and
\IEEEauthorblockN{Taylor Berg-Kirkpatrick}
\IEEEauthorblockA{\textit{Computer Science \& Engineering} \\
\textit{UC San Diego}\\
La Jolla, USA \\
tberg@ucsd.edu}
\and
\IEEEauthorblockN{Julian McAuley}
\IEEEauthorblockA{\textit{Computer Science \& Engineering} \\
\textit{UC San Diego}\\
La Jolla, USA \\
jmcauley@ucsd.edu}
}

\maketitle

\begin{abstract}

The recent explosion of generative AI-Music systems has raised numerous concerns over data copyright, licensing music from musicians, and the conflict between open-source AI and large prestige companies. Such issues highlight the need for publicly available, copyright-free musical data, in which there is a large shortage, particularly for \emph{symbolic} music data. To alleviate this issue, we present \textbf{PDMX}: a large-scale open-source dataset of over 250K public domain MusicXML scores collected from the score-sharing forum MuseScore, making it the largest available copyright-free symbolic music dataset to our knowledge. PDMX additionally includes a wealth of both tag and user interaction metadata, allowing us to efficiently analyze the dataset and filter for high quality user-generated scores. Given the additional metadata afforded by our data collection process, we conduct multitrack music generation experiments evaluating how different representative subsets of PDMX lead to different behaviors in downstream models, and how user-rating statistics can be used as an effective measure of data quality. Examples can be found at \href{https://pnlong.github.io/PDMX.demo/}{\texttt{https://pnlong.github.io/PDMX.demo/}}.

\end{abstract}

\begin{IEEEkeywords}
symbolic music datasets, symbolic music generation, music copyright
\end{IEEEkeywords}

\section{Introduction}\label{introduction}

There has been a recent explosion in development of generative music systems, 
both for symbolic- and audio-domain music generation \cite{forsgren2022riffusion, stableaudio, dong2023mmt,agostinelli2023musiclm, thickstun2023anticipatory,chen2023musicldm}. As such advances have allowed for significant commercial investment in GenAI products (in music and beyond), debate has increased over the legality of training such systems, and in particular, how such models may replace artists and interact with modern copyright law \cite{majumdar2023facing, bulayenko2022ai, drott2021copyright, chen2024using, sturm2019artificial}, including notable high-profile lawsuits against generative music startups \cite{riaa2024}. Such concerns over replacing artists with audio-domain text-to-song generators have too renewed interest in \emph{symbolic}-domain music processing, which allows for greater artist-in-the-loop interaction \cite{thickstun2023anticipatory}. 

A natural alternative to the intense legal process of licensing music 
is to train on public domain data. Unfortunately, this approach remains complex, as many symbolic music datasets currently fail to adequately vet for copyrighted music \cite{liu2022symphony, wikifonia, wang2020pop909, zeng2021musicbert, ens2021building}, let alone explicitly filter for works in the public domain \cite{raffel2016learning, hawthorne2018enabling, hung2021emopia}. This results in practically no available public domain large-scale symbolic datasets, to our knowledge. Additionally, heterogeneity in current symbolic music  has led to most datasets only supporting MIDI format \cite{rothstein1992midi}.
While MIDI is useful for simple symbolic music modeling tasks, it omits an abundance of extra information in musical scores w.r.t to \emph{notating} symbolic music, such as performance directives (e.g.~dynamic markings, articulations), time-located text, and section boundaries, which are only present in more structured formats like MusicXML. 

To remedy these gaps,
we introduce \textbf{PDMX}: a large-scale dataset of over 250K \textbf{P}ublic \textbf{D}omain \textbf{M}usic\textbf{X}ML files collected from the website MuseScore, making it the largest publicly available, copyright-free MusicXML dataset in existence to our knowledge. PDMX includes genre, tag, description, and popularity metadata for every file.
In order to comprehensively parse the MusicXML files present in PDMX, we release an extension of the popular symbolic music processing library MusPy \cite{dong2020muspy}, which we term \emph{MusicRender}, that allows for both storing and accurately realizing common performance cues (e.g.~articulations, dynamic text) into the music itself to support downstream modeling tasks \cite{vaswani2017attention}.

As a test bed, we use an unconditional multitrack music generation task to understand how metadata-based data filtering can affect downstream modeling, highlighting that the user rating data in PDMX can be used as a powerful measure for data quality and enable high quality subsets for training and fine-tuning generative models.
PDMX is released under a CC-BY license as a \emph{fully commercially viable} dataset for symbolic music processing and available at our demo\footnote{\href{https://pnlong.github.io/PDMX.demo/}{\texttt{https://pnlong.github.io/PDMX.demo}}}, as well as all related code for \emph{MusicRender} and trained models.

\section{Related Works}\label{related}

\begin{table*}[h]
\begin{center}
\caption{Summary of Existing Datasets. PDMX is the largest MusicXML dataset, competitive with existing large scale MIDI datasets, and the largest dataset of public domain data.}
\footnotesize
\begin{tabular}{lccccc}
    \toprule
    \textbf{Dataset} & \textbf{Format} & \textbf{Hours} & \textbf{Size} & \textbf{Multitrack} & \textbf{Dataset License Composition} \\
    \midrule
    \RaggedRight{Lakh MIDI (LMD) \cite{raffel2016learning}} & MIDI & $>$9,000 & 174,533 & \cmark & CC-BY 4.0$^\dagger$ \\
    \RaggedRight{SymphonyNet \cite{liu2022symphony}} & MIDI & $>$3,200 & 46,359 & \cmark &  \\
    \RaggedRight{MAESTRO \cite{hawthorne2018enabling}} & MIDI & 201.21 & 1,282 & & CC BY-NC-SA 4.0 \\
    \RaggedRight{Wikifonia Lead Sheet Dataset \cite{wikifonia}} & MusicXML & 198.40 & 6,405 & &  \\
    \RaggedRight{POP909 \cite{wang2020pop909}} & MIDI & 60.00 & 909 & & \\
    \RaggedRight{EMOPIA \cite{hung2021emopia}} & MIDI & 11.0 & 1,078 & & CC BY-NC-SA 4.0 \\
    \RaggedRight{MMD \cite{zeng2021musicbert}} & MIDI & - & 1,524,557 & &  \\
    \RaggedRight{MetaMidi \cite{ens2021building}} & MIDI & - & 612,088 & \cmark &  \\
    \midrule
    \RaggedRight{PDMX} & MusicXML & 6,250 & 254,077 & \cmark & CC-0 / Public Domain \\
    \bottomrule
    \multicolumn{6}{l}{ } \\
    \multicolumn{6}{l}{$\dagger$ While LMD is released under CC-BY 4.0, it has been publically documented that many copyrighted works exist in the dataset \cite{thickstun2023anticipatory}.} \\
\end{tabular}\label{tab:related}
\end{center}
\vspace{-.5cm}
\end{table*}

There exist a number of publicly available datasets for modeling symbolic music (see Tab.~\ref{tab:related}), with a wide variety of licensing scenarios and sizes. 
Most datasets, such as the moderate-sized SymphonyNet  \cite{liu2022symphony}, the large MMD  \cite{zeng2021musicbert} and MetaMidi \cite{ens2021building}, and the smaller Wikifonia \cite{wikifonia} and POP909 \cite{wang2020pop909} datasets offer no clear licensing information, being thus unsafe for copyright issues. While some existing datasets do have explicit licenses, these are either specifically \emph{non-commercial}, such as for MAESTRO \cite{hawthorne2018enabling} and 
EMOPIA\cite{hung2021emopia}, or have documented licensing violations as in LMD \cite{raffel2016learning}.
Additionally, few large datasets contain diverse \emph{multitrack} music, and fewer still present data in the MusicXML format, thus ommiting the notational information present in MusicXML but unsupported in datatypes like MIDI.
Unlike past datasets, PDMX is a large-scale set comprised entirely of CC-0 / Public Domain MusicXML files, containing diverse multitrack symbolic music with song metadata.

\section{Dataset}\label{dataset}

PDMX was created using data scraped from the online score-sharing
forum
MuseScore\footnote{\href{https://musescore.com/}{\texttt{https://musescore.com}}} \cite{xu2024generating}, 
where community members (as well as formal music publishers) can upload their own sheet music arrangements and compositions as MusicXML files, including both licensed and public domain content.
Besides the direct MusicXML files, we additionally scraped the metadata for each score in order to provide a cohesive understanding of the dataset, which includes tag-based descriptions of the score (i.e.~composer, genre), community content in the form of ratings and comments, and most importantly the \emph{license} for each score. 
We then selected for songs with specifically Public Domain Mark (i.e.~already in the public domain prior to uploading) or CC-0 (i.e.~explicitly released into the public domain) licenses. In total, our initial scrape yielded 254,077 MusicXML files, the largest dataset of MusicXML-like data to our knowledge, with a total 6,250.37 hours of music. 

Given the metadata extracted when creating PDMX, we then focused on how such metadata could help us filter the full large-scale dataset into more useful subsets for downstream generative modeling. In particular, we focused
on four main subsets of the data:
\begin{enumerate}
    \item \textbf{All (A)}: The entirety of PDMX.
    \item \textbf{Deduplicated (D)}: The collection of each song's best (in terms of rating if available) unique arrangements.
    \item \textbf{Rated (R)}: Songs with non-zero ratings.
    \item \textbf{Rated and Deduplicated (R$\cap$D)}: The intersection of the previous two subsets.
\end{enumerate}
First, we introduce our \emph{MusicRender} framework for dataset parsing, and then detail how each subset was created.
\begin{table}[t]
\begin{center}
\caption{Subsets of PDMX, including filtering based on deduplication and rating information.}
\scriptsize\begin{tabular}{l|c|ccc}
    \toprule
    \textbf{Subset} & \textbf{Hours} / \textbf{Size} & \textbf{PCE} & \textbf{SC} (\%) & \textbf{GC} (\%) \\
    \midrule
    \RaggedRight{\textbf{A}} & 6,250 / 254K & 2.69 $\pm$ 0.00 & 97.12 $\pm$ 0.01 & 93.75 $\pm$ 0.01 \\
    \RaggedRight{\textbf{D}} & 3,756 / 102K & 2.77 $\pm$ 0.00 & 95.15 $\pm$ 0.02 & 93.79 $\pm$ 0.01 \\
    \RaggedRight{\textbf{R}} & 1,001 / 14K & 2.91 $\pm$ 0.00 & 91.59 $\pm$ 0.07 & 92.59 $\pm$ 0.05 \\
    \RaggedRight{\textbf{R$\cap$D}} & 941 / 13K & 2.90 $\pm$ 0.00 & 91.70 $\pm$ 0.07 & 92.65 $\pm$ 0.05 \\
    \midrule
    \RaggedRight{{Fine-Tuning}} & 595 / 6K & 2.95 $\pm$ 0.00 & 90.54 $\pm$ 0.11 & 92.39 $\pm$ 0.07 \\
    \bottomrule
\end{tabular}\label{tab:subsets}
\end{center}
\vspace{-0.4cm}
\end{table}

\subsection{Parsing MuseScore Data}

Unlike MIDI, MusicXML files are meant for rendering sheet music \emph{as it should be read by a musician} (rather than how it directly sounds), thus containing a wealth of notational information that is common in western music scores.
Existing software capable of parsing these files, namely the Python library MusPy \cite{dong2020muspy}, is not optimized for the task. Given a piece of symbolic music, the resulting MusPy \emph{Music} primarily extracts the note value information, as well as limited metadata like time/key signatures and tempo markings.
However, MusPy fails to parse many MusicXML-specific score objects, including performance directives (e.g.~dynamics, articulations), lyrics, and phrase boundaries, stripping out such content from the notes themselves into a catchall ``annotation" structure.

We design our own data structure, \emph{MusicRender}, as an extension of the MusPy \emph{Music} object to specifically account for these flaws. \emph{MusicRender} supports expanded functionality for parsing MusicXML-specific score attributes using multiple of distinct Python objects for each type of non-note score object (such as rehearsal marks or dynamic text). In particular, MusicRender uses such annotations, along with a number of heuristics, to parse MusicXML files such that its symbolic domain outputs reflect the real perceptual rendering of the notes (i.e.~how notes should be \emph{performed}).
For instance, note durations for a slurred section extend to the next note, and accented notes are louder. Tempo-related annotations, like \emph{ritardandos}, and volume-related features, like \emph{crescendos}, are realized through changes in tempo and velocity.

Similar to MusPy, a \emph{MusicRender} object supports full data I/O to a number of file types (JSON, MIDI), MIR modeling representations (e.g.~piano-roll, music21 \cite{cuthbert2010music21}), and programmatic audio synthesis via Fluidsynth. We thus store PDMX as \emph{MusicRender} JSON files, which are readily loaded into Python environments as \emph{MusicRender} objects with \emph{no} information loss.

\subsection{Data Quality}
\begin{figure}[]
\centerline{\includegraphics[width=\columnwidth, trim={0cm 0cm 0cm 0cm},clip]{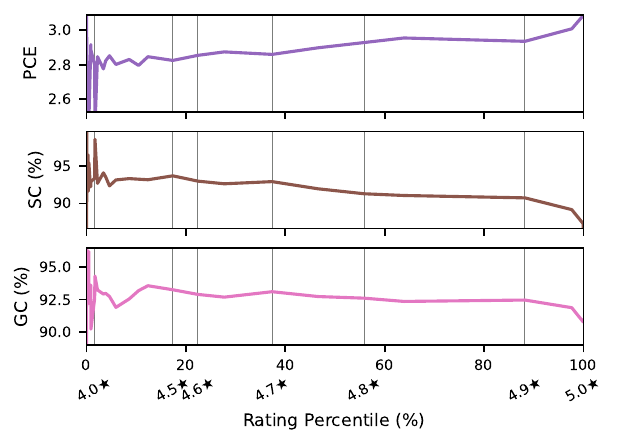}}
\vspace{-0.3cm}
\caption{Pitch Class Entropy (PCE), Scale Consistency (SC), Groove Consistency (GC) vs. rating percentile in PDMX (specific ratings shown as vertical lines). 
Higher-rated songs seem to be more harmonically dynamic (higher PCE, lower SC), yet rating has little effect on rhythm.}
\label{fig:rating}
\vspace{-0.3cm}
\end{figure}

A past problem in existing symbolic music datasets is overall dataset quality, as very limited works exist to assess the ``quality" of symbolic music, and existing high-quality datasets are much smaller \cite{wang2020pop909}.
Due to the crowd-sourced nature of the MuseScore platform, we cannot assume that all songs in the base PDMX dataset are of high quality. While we could use various proxies for quality such as number of views or whether a song has a paywall, we can instead harness MuseScore's crowd-sourcing to our benefit. MuseScore allows individual users to rate a song out of five stars, accessible through the metadata’s ``rating'' attribute. We can use a song's five-star rating average as our metric for data quality.

We refer to the 14,182 rated songs (5.6\% of all songs) as the rated subset (with a rating of zero indicating that a song is unrated).
In practice, ratings actually range from 2.83 to 4.98 stars, 
In Tab.~\ref{tab:subsets}, where we display the Pitch Class Entropy (PCE), Scale Consistency (SC), and Groove Consistency (GC) (following \cite{dong2023mmt,wu2020jazz, thickstun2023anticipatory}), we see that rated songs display more harmonic variation (higher PCE, lower SC) than their unrated counterparts, though a song's rating appears to have little effect on rhythm. Additionally, in Fig.~\ref{fig:rating}, we show these metrics as a function of the rating percentile, displaying that this trend of increasing PCE/decreasing SC holds as the rating increases.

\subsection{Deduplication}

Deduplication is an important part of dataset construction, as heavy duplication within a dataset may bias the dataset towards particular high frequency data points and thus degrade downstream modeling tasks \cite{lee2021deduplicating}.
However, unlike in more traditional domains like text processing \cite{pandey2021prevalent}, string matching techniques using the textual metadata are insufficient for detecting duplicates in PDMX. This is because
many of the same pieces have different titles, such as
``Pachelbel’s Canon in D'' and ``Canon by Pachelbel'', despite being the same piece. Additionally, the notion of ``duplicate" is ill-defined in the context of sheet music, as even two scores of the same song may be arranged differently and for distinct instrumentations.

To address these issue, we first use a pre-trained text embedding model, Sentence-BERT \cite{reimers2019sentence}, to encode song titles as fixed dimension embedding vectors. Formally, we embed a song ``descriptor'' -- a combination of a song's title, subtitle (if applicable), artist, and composer (if different from the artist) -- to combat the scenario where different artists compose works of the same name. We then use cosine similarity to compare embedding vectors and scale those values between zero and one to obtain similarity scores. We set a duplicate threshold of 80\%; that is, we cluster a given song with all other songs $\geq$80\% similar. This threshold was manually verified to capture a reasonable range of duplicate songs. However, deduplicating by song descriptor alone ignores a variety of possible different arrangements, all valuable from a music generation perspective.

For each cluster of duplicate songs, we next group by instrumentation. For example, solo piano arrangements of ``Pachelbel's Canon'' fall into separate clusters from their string quartet counterparts. However, within a single instrumentation cluster can exist multiple unique arrangements (e.g. a beginner versus advanced version). To address this, we use a simple heuristic of total note count to cluster songs (as note count is an easy proxy for large differences in song arrangement). We set a uniqueness threshold of 5\% (i.e.~songs with a $>$5\% difference in note count fall into different clusters). Like the duplicate threshold, we obtained this value through trial and error, feeling that 5\% well-captured overly similar arrangements. Within each similar note count cluster, we finally select the ``best'' arrangement of a song by considering each score's rating and, in the event of a tie, number of notes (the more notes, the better). Therefore, while a score could have tens of duplicates when considering song descriptor alone, within this grouping can exist many unique arrangements.

Using this deduplication strategy, we remove 151,442 songs (nearly 60\% of the base dataset), leaving 102,635 unique arrangements remaining as part of the deduplicated subset.
Notably, while around 85\% of all songs have no duplicates, over 95\% of \emph{rated} songs are unique.
Unlike rating, deduplication appears to have little effect on either harmonic or rhythmic variation.

\subsection{Analysis}

\begin{figure}[]
\centerline{\includegraphics[width=\columnwidth, trim={0cm 0cm 0cm 0cm},clip]{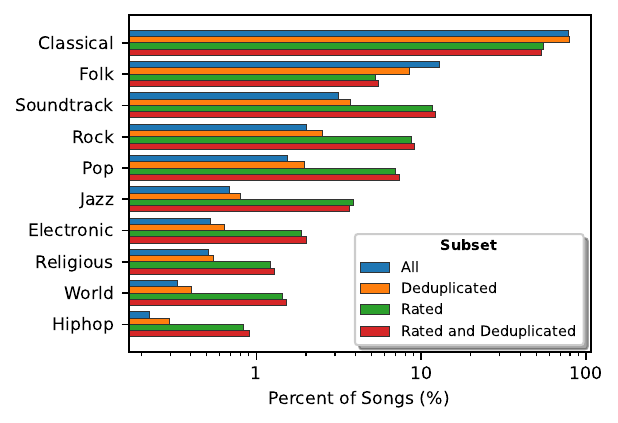}}\vspace{-0.3cm}
\caption{Top-10 Genre distribution in PDMX. 
67\% of songs lack a genre tag. All genres besides the two most common, classical and folk music, display notably higher frequencies in the rated subsets. 
}
\label{fig:genres}
\vspace{-0.4cm}
\end{figure}

As PDMX is a \emph{multitrack} dataset, we first analyze the density of tracks (i.e.~single instrument parts) within a given song. 
Over 90\% of songs in PDMX contain less than five tracks, while over half the dataset consists of solo works. We note that only $\approx$3\% of songs in PDMX have more than five tracks, possibly due to larger multitrack scores (common in orchestral or marching band repertoires) not in the public domain.
PDMX encompasses 20 different genres, the most common being classical and folk music (see Fig.~\ref{fig:genres}). More modern genres, like hip-hop and electronic music, are comparatively much fewer in number,  likely due to limited public domain content for more recent works. Although genre tags are absent from 67\% of songs, only 50\% of \emph{notes} lack a genre, suggesting that genre-labeled works are more dense than unlabelled ones. Additionally, when breaking down the genre distribution by subset (see Fig.~\ref{fig:genres}), we find that the rated (and subsequently rated and deduplicated) subsets of PDMX contain a significantly longer ``tail" of genres than the full dataset (with $\approx$40\% coming from non classical/folk genres), denoting a large amount of unrated \emph{classical} music present in PDMX. 

\subsection{Additional Features}

\begin{figure*}[ht!]
\centerline{\includegraphics[width=\textwidth, trim={0cm 0cm 0cm 0cm},clip]{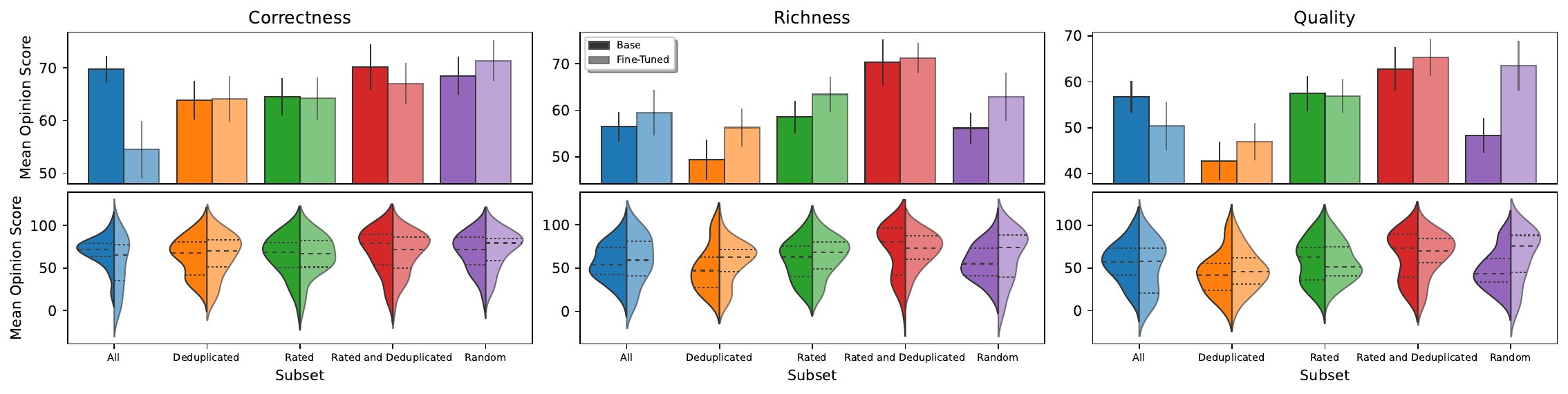}}\vspace{-0.3cm}
\caption{Results from the subjective listening test. For each subset and rating axis, the mean opinion scores of the base and fine-tuned models are displayed on the left and right, respectively. The \textbf{R$\cap$D} (red) subset generally performs the best, and fine-tuning on $>$50\% rated data improves richness and quality.}
\label{fig:listening}
\end{figure*}

While we have focused on highlighting PDMX’s note-based content and parts of its metadata,
we note that PDMX includes a number of extra features. Regarding metadata, PDMX includes rich rating and user comment information, enabling work on symbolic music recommendation \cite{chen2005music, lisena2022midi2vec} and preference modeling. Notably, as PDMX is a MusicXML dataset, it includes 12M time-aligned performance directives such as tempo text (e.g.~\emph{adagio}), dynamic hairpins (e.g.~\emph{crescendo}), note articulations (e.g.~\emph{staccato}), and section markings, which are parsed by the \emph{MusicRender} framework and may be used both for discriminative tagging tasks, expressive music rendering, and even as conditions for controllable music generation \cite{thickstun2023anticipatory, Novack2024Ditto}. PDMX also includes over 10M lyric tokens, opening the door for further research on lyric-to-score and score-to-lyric tasks \cite{ding2024songcomposer, sheng2021songmass, ogawa2021tohoku}. We plan to fully release PDMX as 44.1kHz synthetic audio rendered through Fluidsynth, creating a large corpus for music transcription \cite{gardner2021mt3} and audio-domain music generation \cite{chen2023musicldm, Novack2024Ditto}.

\section{Experiments}\label{experiments}

\begin{table}[]
\vspace{-0.5cm}
\begin{center}
\caption{Quantitative Results. For all metrics, performance is determined by closeness to the fine-tuning subset's statistics.}
\scriptsize\begin{tabular}{lc|ccc}
    \toprule
    \textbf{Subset} & \textbf{Fine-Tuned} & \textbf{PCE} & \textbf{SC} (\%) & \textbf{GC} (\%) \\
    \midrule
    \RaggedRight{\textbf{A}} &  & 2.66 $\pm$ 0.01 & 97.65 $\pm$ 0.13 & 93.70 $\pm$ 0.11 \\
    \RaggedRight{\textbf{D}} &  & 2.66 $\pm$ 0.01 & 96.32 $\pm$ 0.16 & 93.83 $\pm$ 0.15 \\
    \RaggedRight{\textbf{R}} &  & \bf{2.80 $\pm$ 0.01} & \underline{93.89 $\pm$ 0.20} & \underline{91.83 $\pm$ 0.22} \\
    \RaggedRight{\textbf{R$\cap$D}} &  & \underline{2.80 $\pm$ 0.01} & \bf{93.32 $\pm$ 0.21} & \bf{91.90 $\pm$ 0.21} \\
    \RaggedRight{\textbf{Random}} &  & 2.68 $\pm$ 0.01 & 97.57 $\pm$ 0.12 & 93.46 $\pm$ 0.12 \\
    \midrule
    \RaggedRight{\textbf{A}} & \cmark & 2.81 $\pm$ 0.01 & \bf{92.42 $\pm$ 0.23} & 90.75 $\pm$ 0.25 \\
    \RaggedRight{\textbf{D}} & \cmark & 2.79 $\pm$ 0.01 & 93.17 $\pm$ 0.22 & \bf{90.93 $\pm$ 0.24} \\
    \RaggedRight{\textbf{R}} & \cmark & \underline{2.84 $\pm$ 0.01} & 92.77 $\pm$ 0.22 & 90.68 $\pm$ 0.24 \\
    \RaggedRight{\textbf{R$\cap$D}} & \cmark & \bf{2.85 $\pm$ 0.01} & \underline{92.51 $\pm$ 0.22} & 90.38 $\pm$ 0.26 \\
    \RaggedRight{\textbf{Random}} & \cmark & 2.79 $\pm$ 0.01 & 93.37 $\pm$ 0.21 & \underline{90.83 $\pm$ 0.27} \\
    \bottomrule
\end{tabular}\label{tab:results}
\end{center}
\vspace{-0.6cm}
\end{table}

The metadata included in PDMX presents an opportunity to analyze how data quality filtering and deduplication affects downstream symbolic music modeling.
Specifically, viewing each score in PDMX as a sequence of notes $\bm{n} = \bm{n}_1, \dots, \bm{n}_N$, our goal is to learn an autoregressive model $p_{\theta}(\bm{n}_i \mid \bm{n}_{1:i-1})$ that we can generate symbolic music from. We focus on our four main subsets of PDMX (detailed in Tab.~\ref{tab:subsets}), as well as a \textbf{Random} subset of songs sampled from the full dataset at the size of the rated and deduplicated subset.
We additionally reserve the top 50\% (in terms of rating, specifically $>$4.74 stars) from the rated and deduplicated subset as a potentially ``high quality" subset for fine-tuning.
In this setup, we seek to answer two main questions: (1) How do data quality and deduplication interact to determine symbolic modeling results?
and (2) Does fine-tuning on small but high quality data meaningfully change behavior?

\vspace{-0.1cm}
\subsection{Experimental Setup and Metrics}

In this work, we use the tokenization scheme REMI+ \cite{von2022figaro}, an extension to the REMI input representation that allows for multitrack music \cite{huang2020pop}. Namely, we represent a song as a sequence of notes $\bm{n}_{i}$, where each note is represented by a series of five variables: beat, position, pitch, duration, instrument. We employ a metrical timing system (i.e.~time is represented in beats rather than seconds).

For all experiments, we use a REMI+ style decoder-only transformer, with 6 layers, 8 attention heads, and a hidden dimension of 512, totaling at $\approx$20M parameters. We use an absolute positional embedding with a maximum sequence length of 1024. We train each model on a single A6000 GPU for 100K steps with a batch size of 12, learning rate of 5e-4, and the Adam optimizer with default hyperparameters. For fine-tuning, we employ a smaller learning rate of 5e-5, and only train for 5K steps. 
Following past work \cite{dong2023mmt, wu2020jazz, thickstun2023anticipatory}, we report the PCE, SC, and GC across 1,200 generations per model, which measure how well the model captures the underlying musical patterns of the data.

\subsection{Results}

\subsubsection{Objective Metrics}

In Tab.~\ref{tab:results}, we observe that between the five base models, those trained on the rated subsets display greater harmonic and rhythmic diversity than those not, and have the closest statistics to the fine-tuning subset. However, once all models fine-tune on the $50\%$ best rated scores, this distinction goes away, with all five fine-tuned models showing more similar metrics.

\subsubsection{Subjective Listening Test}

To measure music quality across our ten models (five base, five fine-tuned), we conducted a listening test with 12 participants. In the questionnaire, each participant listened to 30 different samples randomly chosen from a pool of 10 samples per model. Following past work \cite{wu2023compose}, for each sample, users were asked to rate the generation (from 0 to 100) along three axes:
\begin{itemize}
    \item \textbf{Correctness:} Is the music free of inharmonious notes, unnatural rhythms, and awkward phrasing?
    \item \textbf{Richness:} Is the sample musically / harmonically interesting?
    \item \textbf{Quality:} Subjectively, how much do you like the generation?
\end{itemize}
In Fig.~\ref{fig:listening}, we show the average values (top) and violin plots (bottom) for each model along each rating axis. In comparing different modeling subsets (ignoring fine-tuning), we find that correctness is consistent across all subsets. For richness and quality however, the \textbf{R$\cap$D} subset (i.e.~our most filtered subset, in red) consistently shows the highest scores, followed by \textbf{R} (green), \textbf{A} (blue), \textbf{Random} (purple), and \textbf{D} (orange), suggesting that our deduplication strategy is primarily useful for rated songs (as we pick the highest rated duplicate) rather than unrated (which is chosen at random).

Regarding fine-tuning, we find that this process
increases richness in \emph{all} models, and improves quality in three (\textbf{D}, \textbf{R$\cap$D}, \textbf{Random}), with a particularly strong effect on the Random (purple) model, converting it into our second-highest rated in quality overall. There is also a noticeable negative effect on correctness in the All (blue) model, suggesting that the model may have overfit to overly simple examples in the full dataset. These results together show the strength of our deduplication and filtering strategies as a way to perform significant dataset distillation (as the \textbf{R$\cap$D} model performs best despite seeing only 15\% of the data).

\section{Conclusion}\label{conclusion}

We present \textbf{PDMX}: the largest dataset of public domain MusicXML files 
to our knowledge. Using our proposed \emph{MusicRender} package for efficient parsing and hand-crafted filtering algorithms, we show promising results for unconditional multitrack generation that indicate improved performance when filtering for high-quality scores. In the future, we hope to investigate ways to use the large dataset as an effective pretraining mechanism for symbolic generation models, and  
utilize PDMX's wealth of extra performance directive annotations for 
fine-grained time-located text-to-music generation. We also plan to explore PDMX's capacity for discriminative MIR tasks and uses involving its large pool of lyric data.

\section{Acknowledgements}\label{acknowledgements}

We thank Hao-Wen Dong for his efforts in scraping MuseScore and compiling the data used to create PDMX.

\bibliographystyle{IEEEtran}
\bibliography{references}

\newpage

\end{document}